\documentclass[12pt]{iopart}

\usepackage{iopams}
\usepackage{setstack}
\usepackage{graphicx}
\usepackage{bm}
\usepackage{hyperref}
\usepackage{verbatim}
\hypersetup{colorlinks=false}

\begin{document}

\title[Confluent Heun functions and the physics of black holes]{Confluent Heun functions and the physics of black holes: resonant frequencies, Hawking radiation and scattering of scalar waves}

\author{H. S. Vieira$^{1,2}$ and V. B. Bezerra$^{1}$}

\address{$^{1}$ Departamento de F\'{i}sica, Universidade Federal da Para\'{i}ba, Caixa Postal 5008, CEP 58051-970, Jo\~{a}o Pessoa, PB, Brazil}
\address{$^{2}$ Centro de Ci\^{e}ncias, Tecnologia e Sa\'{u}de, Universidade Estadual da Para\'{i}ba, CEP 58233-000, Araruna, PB, Brazil}

\ead{horacio.santana.vieira@hotmail.com and valdir@fisica.ufpb.br}

\begin{abstract}
We apply the confluent Heun functions to study the resonant frequencies (quasispectrum), the Hawking radiation and the scattering process of scalar waves, in a class of spacetimes, namely, the ones generated by a Kerr-Newman-Kasuya spacetime (dyon black hole) and a Reissner-Nordstr\"{o}m black hole surrounded by a magnetic field (Ernst spacetime). In both spacetimes, the solutions for the angular and radial parts of the corresponding Klein-Gordon equations are obtained exactly, for massive and massless fields, respectively. The special cases of Kerr and Schwarzschild black holes are analyzed and the solutions obtained, as well as in the case of a Schwarzschild black hole surrounded by a magnetic field. In all these special situations, the resonant frequencies, Hawking radiation and scattering are studied.
\end{abstract}

%\pacs{02.30.Gp, 04.20.Jb, 04.70.-s, 04.80.Cc, 47.35.Rs, 47.90.+a}

%\keywords{massless Klein-Gordon equation, rotating acoustic black hole, canonical acoustic black hole, confluent Heun functions, dumb body radiation}

%\msc{81Q05, 83C45, 83C57, 83C75}

%\preprint{AIP/123-QED}

\maketitle

%\begin{quotation}
%...
%\end{quotation}

%
%%%%%%%%%%%%%%%%%%%%%%%%%%%%%%%%%%%%%%%%%%%%%%%%%%%%%%%%%%%%%%%%%%%%%%%%%%%%%%%%%%%%%%%%%%%%%% Introduction
%
\section{Introduction}
The theoretical studies concerning physical processes which occur in the spacetime surrounding black holes are of special interest and, certainly, can help us to understand the physics of these interesting objects predicted by general relativity. Among these studies, we can mention the ones corresponding to resonant frequencies \cite{LettNuovoCimento.15.257,PhysRevD.22.2323,ZhEkspTeorFiz.92.369}, scatterring of particles and fields of different spins \cite{PhysRevD.16.937,PhysRevD.18.1030,PhysRevD.75.104012,IntJModPhysD.21.1250045,CommunTheorPhys.62.227} and Hawking radiation \cite{CommunMathPhys.43.199}.

During last years, the Heun functions have gained more and more importance due to the large number of applications in different areas of physics (see \cite{arXiv:1101.0471v8} and references therein) and in special, in the solutions of problems related to a scalar field in gravitational backgrounds \cite{JPhysAMathGen.9.1631}. The use of the Heun's functions permits us, for example, to find the exact solutions of the Klein-Gordon equation in some black hole spacetimes \cite{ClassQuantumGrav.31.045003}. Otherwise, without the use of these functions, the scalar solutions, for massless and massive scalar fields, will be possible only for some specific regions very close and far away from the black hole horizons.

In the present paper, we apply the confluent Heun functions to obtain the solutions of the Klein-Gordon equations for a charged massive scalar field in the Kerr-Newman-Kasuya spacetime (dyon black hole), and for a massless scalar field in the Reissner-Nordstr\"{o}m black hole surrounded by a magnetic field (Ernst spacetime). These solutions, which are given in terms of the confluent Heun functions, are used to examine the resonant frequencies, the Hawking radiation and the scattering process of scalar waves.

This paper is organized as follows. In Section 2, we present the solutions of the Klein-Gordon equation for a charged massive scalar field in the dyon black hole spacetime, for both angular and radial parts. We also obtain the resonant frequencies. We discuss the scattering process and present a result obtained in the literature \cite{AnnPhys.362.576}, as a brief review, on the Hawking radiation with the aim to cover the three different topics for each configuration spacetimes considered, and we present and discuss some special cases. In Section 3, we present the solutions for the both angular and radial parts of the Klein-Gordon equation, for a massless scalar field, in the Reissner-Nordstr\"{o}m black hole surrounded by a magnetic field. We obtain the resonant frequencies for this case. We also discuss the scattering and the Hawking radiation of massless scalar waves, and we consider the special case that corresponds to the Schwarzschild black hole surrounded by a magnetic field. Finally, in Section 4, the conclusions are given.
%
%%%%%%%%%%%%%%%%%%%%%%%%%%%%%%%%%%%%%%%%%%%%%%%%%%%%%%%%%%%%%%%%%%%%%%%%%%%%%%%%%%%%%%%%%%%%%% Equation of motion for charged massive scalar fields
%
\section{Charged massive scalar field equation in the Kerr-Newman-Kasuya spacetime}
The covariant Klein-Gordon equation that describes the behavior of charged scalar field in a general curved spacetime interacting with an electromagnetic field is given by
\begin{eqnarray}
& & \left[\frac{1}{\sqrt{-g}}\partial_{\sigma}\left(g^{\sigma\tau}\sqrt{-g}\partial_{\tau}\right)-ie(\partial_{\sigma}A^{\sigma})-2ieA^{\sigma}\partial_{\sigma}\right.\nonumber\\
& & -\left.\frac{ie}{\sqrt{-g}}A^{\sigma}(\partial_{\sigma}\sqrt{-g})-e^{2}A^{\sigma}A_{\sigma}-\mu_{0}^{2}\right]\Psi=0\ ,
\label{eq:Klein-Gordon_gauge}
\end{eqnarray}
where $\Psi$ is the complex scalar field, $\mu_{0}$ is the mass parameter, $e$ is the charge of the particle, and $A^{\sigma}$ is the 4-vector of the electromagnetic field.

Now, let us consider the curved spacetime generated by a dyon black hole, whose line element in the Boyer-Lindquist coordinates is given by \cite{PhysRevD.25.995}
\begin{eqnarray}
ds^{2} & = & g_{\sigma\tau}dx^{\sigma}dx^{\tau}\nonumber\\
 & = & -\frac{\Delta}{\rho^{2}}(dt-a\sin^{2}\theta\ d\phi)^{2}+\frac{\rho^2}{\Delta}\ dr^{2}+\rho^{2}\ d\theta^{2}\nonumber\\
 & & +\frac{\sin^2\theta}{\rho^2}[(r^2+a^2)d\phi-a\ dt]^{2}\ ,
\label{eq:metrica_Kerr-Newman-Kasuya}
\end{eqnarray}
while the vector potential is such that
\begin{eqnarray}
A_{\sigma}dx^{\sigma} & = & -\left(Q_{e}\frac{r}{\rho^{2}}-Q_{m}\frac{a\cos\theta}{\rho^{2}}\right)dt\nonumber\\
& & +\left[Q_{e}\frac{ar\sin^{2}\theta}{\rho^{2}}+Q_{m}\left(\xi-\frac{r^{2}+a^{2}}{\rho^{2}}\cos\theta\right)\right]d\phi\ ,
\label{eq:potencial_EM_Kerr-Newman-Kasuya}
\end{eqnarray}
with
\begin{equation}
\Delta=r^{2}-2Mr+a^{2}+Q_{e}^{2}+Q_{m}^{2}=(r-r_{+})(r-r_{-})\ ,
\label{eq:Delta_metrica_Kerr-Newman-Kasuya}
\end{equation}
\begin{equation}
r_{\pm}=M \pm [M^{2}-(a^{2}+Q_{e}^{2}+Q_{m}^{2})]^{\frac{1}{2}}\ ,
\label{eq:sol_padrao_Kerr-Newman-Kasuya}
\end{equation}
\begin{equation}
\rho^{2}=r^{2}+a^{2}\cos^{2}\theta\ ,
\label{eq:rho_metrica_Kerr-Newman-Kasuya}
\end{equation}
where $a=J/M$ is the angular momentum per mass, $Q_{e}$ is the electric charge, $Q_{m}$ is the magnetic charge, $M$ is the mass (energy), and $\xi=\pm 1$ (these two signs correspond to the two gauges of this background). Note that the units $G \equiv c \equiv \hbar \equiv 1$ were chosen.

Due to the time independence and symmetry of the spacetime, we can assume that the wave function can be written as
\begin{equation}
\Psi=\Psi(\mathbf{r},t)=R(r)S(\theta)\mbox{e}^{im\phi}\mbox{e}^{-i\omega t}\ ,
\label{eq:separacao_variaveis_Kerr-Newman-Kasuya}
\end{equation}
where $m=\pm 1,\pm 2,\pm 3,\ldots$ is the azimuthal quantum number, and $\omega$ is the frequency (energy in the units used).

Thus, the exact solution of the angular part of the Klein-Gordon equation, $S(\theta)$, for a charged massive scalar field in the dyon black hole spacetime, over the entire range $0 \leq z < \infty$, was obtained recently \cite{AnnPhys.362.576}, which is given by
\begin{eqnarray}
S(z) & = & \mbox{e}^{\frac{1}{2}\alpha z}z^{\frac{1}{2}\beta}(z-1)^{\frac{1}{2}\gamma}\{C_{1}\ \mbox{HeunC}(\alpha,\beta,\gamma,\delta,\eta;z)\nonumber\\
& & +C_{2}\ z^{-\beta}\ \mbox{HeunC}(\alpha,-\beta,\gamma,\delta,\eta;z)\}\ ,
\label{eq:solucao_geral_angular_Kerr-Newman-Kasuya_case2_z}
\end{eqnarray}
with
\begin{equation}
z=\frac{\cos\theta+1}{2}\ ,
\label{eq:coord_angular_Kerr-Newman-Kasuya_case2_z}
\end{equation}
where $C_{1}$ and $C_{2}$ are constants, $\mbox{HeunC}(\alpha,\pm\beta,\gamma,\delta,\eta;z)$ are the confluent Heun functions \cite{Slavyanov:2000,ClassQuantumGrav.27.135001}, and the parameters $\alpha$, $\beta$, $\gamma$, $\delta$, and $\eta$ are expressed as
\begin{equation}
\alpha=4 a (\omega ^2-\mu_{0}^{2})^{\frac{1}{2}}\ ,
\label{eq:alpha_angular_Kerr-Newman-Kasuya_case2_z}
\end{equation}
\begin{equation}
\beta=m-e Q_{m}(\xi +1)\ ,
\label{eq:beta_angular_Kerr-Newman-Kasuya_case2_z}
\end{equation}
\begin{equation}
\gamma=m-e Q_{m}(\xi -1)\ ,
\label{eq:gamma_angular_Kerr-Newman-Kasuya_case2_z}
\end{equation}
\begin{equation}
\delta=-4 a \omega e Q_{m}\ ,
\label{eq:delta_angular_Kerr-Newman-Kasuya_case2_z}
\end{equation}
\begin{equation}
\eta=-\frac{-2 a^2 \mu_{0}^{2}+4 a \omega  [m-e Q_{m}(\xi +1)]+e^2 Q_{m}^{2}-(m-e \xi Q_{m})^2+2 \lambda }{2}\ .
\label{eq:eta_angular_Kerr-Newman-Kasuya_case2_z}
\end{equation}
These two functions form linearly independent solutions of the confluent Heun dif\-fer\-en\-tial equation, provided $\beta$ is not integer.

For the radial part of the Klein-Gordon equation, $R(r)$, the exact solution over the entire range $0 \leq x < \infty$, was also obtained recently \cite{AnnPhys.362.576} and is given by
\begin{eqnarray}
R(x) & = & \mbox{e}^{\frac{1}{2}\alpha x}x^{\frac{1}{2}\beta}(x-1)^{\frac{1}{2}\gamma}\{C_{1}\ \mbox{HeunC}(\alpha,\beta,\gamma,\delta,\eta;x)\nonumber\\
& & +C_{2}\ x^{-\beta}\ \mbox{HeunC}(\alpha,-\beta,\gamma,\delta,\eta;x)\}\ ,
\label{eq:solucao_geral_radial_Kerr-Newman-Kasuya_case2_x}
\end{eqnarray}
with
\begin{equation}
x=\frac{r-r_{+}}{r_{-}-r_{+}}\ ,
\label{eq:homog_subs_radial_Kerr-Newman-Kasuya_case2_x}
\end{equation}
where $C_{1}$ and $C_{2}$ are constants, and the parameters $\alpha$, $\beta$, $\gamma$, $\delta$, and $\eta$ are expressed as
\begin{equation}
\alpha=2(\mu_{0} ^2-\omega ^2)^{\frac{1}{2}} (r_{+}-r_{-})\ ,
\label{eq:alpha_radial_Kerr-Newman-Kasuya_case2_x}
\end{equation}
\begin{equation}
\beta=2i\frac{ \omega(r_{+}^2+a^2) -a m - e ( Q_{e} r_{+} - \xi Q_{m} a ) }{ r_{+}-r_{-}}\ ,
\label{eq:beta_radial_Kerr-Newman-Kasuya_case2_x}
\end{equation}
\begin{equation}
\gamma=2i\frac{ \omega(r_{-}^2+a^2) -a m - e ( Q_{e} r_{-} - \xi Q_{m} a ) }{ r_{+}-r_{-}}\ ,
\label{eq:gamma_radial_Kerr-Newman-Kasuya_case2_x}
\end{equation}
\begin{equation}
\delta=[2 \omega e Q_{e} + (\mu_{0} ^2-2 \omega ^2)(r_{+}+r_{-})](r_{+}-r_{-})\ ,
\label{eq:delta_radial_Kerr-Newman-Kasuya_case2_x}
\end{equation}
\begin{eqnarray}
\eta & = & -\frac{2 a^{2}[a \omega-(m-e \xi Q_{m})]^{2} +4a^2 \omega ^2 r_{+}r_{-}}{(r_{+}-r_{-})^2 }\nonumber\\
& - & \frac{(r_{+}-r_{-})^2 (\lambda +\mu_{0} ^2 r_{+}^2)-4 a \omega  (m- e \xi  Q_{m}) r_{+} r_{-}}{(r_{+}-r_{-})^2 }\nonumber\\
& - & \frac{2 e^2 Q_{e}^2 r_{+} r_{-}+2 e Q_{e} r_{+}^2 	\omega  (r_{+}-3 r_{-})-2 r_{+}^3 \omega ^2 (r_{+}-2 r_{-})}{(r_{+}-r_{-})^2 }\nonumber\\
& - & \frac{2 a e Q_{e} (r_{+}+r_{-}) (m-e \xi  Q_{m})}{(r_{+}-r_{-})^2 }\nonumber\\
& - & \frac{-2 a^2 \omega e Q_{e}  (r_{+}+r_{-})}{(r_{+}-r_{-})^2 }\ .
\label{eq:eta_radial_Kerr-Newman-Kasuya_case2_x}
\end{eqnarray}
As in the case corresponding to the angular solution, in the present situation there is not any specific physical reason to impose that $\beta$ should be integer.

After the brief review of the solutions of the Klein-Gordon equation, let us now, take into account the properties of the confluent Heun functions to discuss the resonant frequencies (quasispectrum), the scattering process and Hawking radiation of scalar waves by a dyonic black hole.
%
%%%%%%%%%%%%%%%%%%%%%%%%%%%%%%%%%%%%%%%%%%%%%%%%%%%%%%%%%%%%%%%%%%%%%%%%%%%%%%%%%%%%%%%%%%%%%% Resonant frequencies
%
\subsection{Resonant frequencies}
In this section we present a new simple technique for investigating the resonant frequencies (quasispectrum) for a charged scalar field in a dyonic black hole background, using directly the exact solutions of the Klein-Gordon equation.

Many authors have considered the quasispectrum of massive fields in several black hole spacetimes. The terminology quasinormal modes has been used by Simone et al. \cite{ClassQuantumGrav.9.963} and Fiziev \cite{ClassQuantumGrav.23.2447}, and bound states by Dolan et al. \cite{ClassQuantumGrav.32.184001}. Here, we use quasispectrum generically, which have no implication with time-reversal symmetry. Quasispectrum is not to be confused with quasinormal modes or bound states. More specifically, quasinormal modes represent the scattering resonances of the fields in the black hole spacetime, where these characteristic resonances correspond to the poles of the (frequency dependent) transmission and reflection amplitudes (or are associated with a maximum in the effective potential), while for computing the discrete spectrum of bound states, the radial solution should be finite on the horizon and decay exponentially towards infinity.

We now formulate a practical method for computing the discrete spectrum of energy levels. In essence, we wish to solve Eq.~(\ref{eq:separacao_variaveis_Kerr-Newman-Kasuya}) for the radial part $R(r)$ subject to certain boundary conditions. The solution should be finite on the horizon and well behaved far from the black hole.

In order to compute these field energies we need to impose boundary conditions on the solutions at the asymptotic region (infinity), which in this case, requires the necessary condition to have a polynomial form for $R(x)$, which is obtained by imposing the so called $\delta$-condition \cite{JPhysAMathTheor.43.035203}
\begin{equation}
\frac{\delta}{\alpha}+\frac{\beta+\gamma}{2}+1=-n\ ,\qquad n=0,1,2,\ldots\ .
\label{eq:cond_polin_1}
\end{equation}

Thus, we obtain the following expression which involves the resonant frequencies
\begin{equation}
\frac{\omega e Q_{e}+M\mu_{0}^{2}-2M\omega^{2}}{(\mu_{0}^{2}-\omega^{2})^{\frac{1}{2}}}+\frac{i}{2}\left[\left(\frac{\omega}{\kappa_{+}}+\frac{\omega}{\kappa_{-}}\right)-\left(\frac{\omega_{+}}{\kappa_{+}}+\frac{\omega_{-}}{\kappa_{-}}\right)\right]=-(n+1)\ ,
\label{eq:quasispectrum_modes_Kerr-Newman-Kasuya_case2_x}
\end{equation}
where
\begin{equation}
\kappa_{\pm}=\frac{1}{2}\frac{r_{\pm}-r_{\mp}}{r_{\pm}^{2}+a^{2}}\ ,
\label{eq:acel_grav_Kerr-Newman-Kasuya}
\end{equation}
\begin{equation}
\omega_{\pm}=m\Omega_{\pm}+e\Phi_{\pm}\ ,
\label{eq:omega_Kerr-Newman-Kasuya}
\end{equation}
\begin{equation}
\Omega_{\pm}=\frac{a}{r_{\pm}^{2}+a^{2}}\ ,
\label{eq:vel_ang_Kerr-Newman-Kasuya}
\end{equation}
\begin{equation}
\Phi_{\pm}=\frac{Q_{e}r_{\pm}-\xi Q_{m}a}{r_{\pm}^{2}+a^{2}}\ .
\label{eq:pot_ele_Kerr-Newman-Kasuya}
\end{equation}

It is not possible to obtain an analytic expression for $\omega_{n}$ from Eq.~(\ref{eq:quasispectrum_modes_Kerr-Newman-Kasuya_case2_x}), however, there are several numerical methods that can be used to obtain approximate expressions for each energy level \cite{PhysRevD.84.127502}.
%
%%%%%%%%%%%%%%%%%%%%%%%%%%%%%%%%%%%%%%%%%%%%%%%%%%%%%%%%%%%%%%%%%%%%%%%%%%%%%%%%%%%%%%%%%%%%%% Massless scalar fields
%
%\subsection{Massless scalar fields}

If we consider massless scalar fields, the frequency corresponding to the quasi\-spectrum, $\omega_{QS}$, can be decomposed into the real, $\omega_{R}$, and imaginary, $\omega_{I}$, parts, and therefore, $\omega_{QS}=\omega_{R}+i\omega_{I}$. In this situation, the quasispectrum of massless scalar fields, $\mu_{0}=0$, is such that
\begin{equation}
\omega_{R}=\frac{2\kappa_{+}\kappa_{-}}{4M\kappa_{+}\kappa_{-}+\kappa_{+}+\kappa_{-}}\left[eQ_{e}+\frac{1}{2}\left(\frac{\omega_{+}}{\kappa_{+}}+\frac{\omega_{-}}{\kappa_{-}}\right)\right]
\label{eq:omegaR_Kerr-Newman-Kasuya_case2_x_massless}
\end{equation}
and
\begin{equation}
\omega_{I}=\frac{2\kappa_{+}\kappa_{-}}{4M\kappa_{+}\kappa_{-}+\kappa_{+}+\kappa_{-}}(n+1)\ ,
\label{eq:omegaI_Kerr-Newman-Kasuya_case2_x_massless}
\end{equation}
with $n \geq 0$.
%
%%%%%%%%%%%%%%%%%%%%%%%%%%%%%%%%%%%%%%%%%%%%%%%%%%%%%%%%%%%%%%%%%%%%%%%%%%%%%%%%%%%%%%%%%%%%%% Scattering
%
\subsection{Scattering}
If we consider the neighborhood of the irregular singular point at infinity, the two solutions of the confluent Heun equation exist, and in general they can be expanded (in a sector) in the following asymptotic series \cite{Ronveaux:1995}
\begin{equation}
\mbox{HeunC}(\alpha,\beta,\gamma,\delta,\eta;x) \sim C_{1}\ x^{-\left(\frac{\beta+\gamma+2}{2}+\frac{\delta}{\alpha}\right)}+C_{2}\ \mbox{e}^{-\alpha x}x^{-\left(\frac{\beta+\gamma+2}{2}-\frac{\delta}{\alpha}\right)}\ ,
\label{eq:assintotica_HeunC_x_grande}
\end{equation}
where we are keeping only the first term of this power-series asymptotics.

Thus, from Eqs.~(\ref{eq:homog_subs_radial_Kerr-Newman-Kasuya_case2_x}) and (\ref{eq:assintotica_HeunC_x_grande}) we can see that the radial solution given by Eq.~(\ref{eq:solucao_geral_radial_Kerr-Newman-Kasuya_case2_x}), far from the black hole, that is, when $|r| \rightarrow \infty$, which implies that $|x| \rightarrow \infty$, behaves asymptotically as
\begin{equation}
R(x) \sim \frac{1}{x}\{C_{1}\ \mbox{e}^{\frac{1}{2}\alpha x}x^{-\frac{\delta}{\alpha}}+C_{2}\ \mbox{e}^{-\frac{1}{2}\alpha x}x^{\frac{\delta}{\alpha}}\}\ ,
\label{eq:scattering_wave_Kerr-Newman-Kasuya_case2_x_1}
\end{equation}
where all constants are included in $C_{1}$ and $C_{2}$. Thus, the radial wave function can be written as
\begin{equation}
R_{\lambda}(x) \sim C_{\lambda}\ \frac{1}{x} \sin\left[-i\frac{1}{2}\alpha x+i\frac{\delta}{\alpha}\ln x+\sigma_{\lambda}(\omega)\right]\ ,
\label{eq:scattering_wave_Kerr-Newman-Kasuya_case2_x_2}
\end{equation}
where $\sigma_{\lambda}(\omega)$ is the phase shift.

From Eqs.~(\ref{eq:alpha_radial_Kerr-Newman-Kasuya_case2_x}) and (\ref{eq:delta_radial_Kerr-Newman-Kasuya_case2_x}), we obtain
\begin{equation}
-i\frac{1}{2}\alpha=(\omega ^2-\mu_{0} ^2)^{\frac{1}{2}} (r_{+}-r_{-})=k_{0} (r_{+}-r_{-})\ ,
\label{eq:-ialpha/2_radial_Kerr-Newman-Kasuya_case2_x}
\end{equation}
\begin{equation}
i\frac{\delta}{\alpha}=\frac{2 \omega e Q_{e} + (\mu_{0} ^2-2 \omega ^2)(r_{+}+r_{-})}{2(\omega ^2-\mu_{0} ^2)^{\frac{1}{2}}}=\gamma_{0}\ ,
\label{eq:idelta/alpha_radial_Kerr-Newman-Kasuya_case2_x}
\end{equation}
where $k_{0}^{2} \equiv \omega ^2-\mu_{0} ^2$ and $\gamma_{0} \equiv (\omega e Q_{e} + M\mu_{0} ^2-2M \omega ^2)/(\omega ^2-\mu_{0} ^2)^{\frac{1}{2}}$.

Therefore, the regular partial wave solution has the asymptotic form
\begin{equation}
R_{\lambda}(r) \sim C_{\lambda}\ \frac{1}{r} \sin[k_{0}r-\gamma_{0}\ln r+\sigma_{\lambda}(\omega)]\ .
\label{eq:sol_scattering_wave_Kerr-Newman-Kasuya_case2_x_2}
\end{equation}

These solutions for the scalar fields far from the black hole will be useful to investigate the scattering of charged massive scalar fields. It is worth calling attention to the fact that we are using the analytical solution of the radial part of the Klein-Gordon equation in the spacetime under consideration.

Indeed, the phase shift, $\sigma_{\lambda}$, is not a simple function of $l$, and thus, the exact expression of the scattering amplitude is not available but just an approximate one is obtained, as is shown for instance by Abramov et al. \cite{JPhysBAtomMolecPhys.12.1761}.
%
%%%%%%%%%%%%%%%%%%%%%%%%%%%%%%%%%%%%%%%%%%%%%%%%%%%%%%%%%%%%%%%%%%%%%%%%%%%%%%%%%%%%%%%%%%%%%% Hawking radiation
%
\subsection{Hawking radiation}
The exact solution of the Hawking effect for charged massive scalar particles in the dyon black hole was obtained recently \cite{AnnPhys.362.576} (if we put $b=1$ into its Eqs.~(93), (94), (102) and (107)). In this subsection we will present briefly the results concerning Hawking radiation \cite{AnnPhys.362.576}, for the sake of completeness, in view of the fact that we will present new results on this subject, in the next section.

The ingoing and outgoing wave solutions, on the black hole exterior horizon surface, are given by
\begin{equation}
\Psi_{in}=\mbox{e}^{-i \omega t}(r-r_{+})^{-\frac{i}{2\kappa_{+}}(\omega-\omega_{0})}\ ,
\label{eq:sol_in_1_Kerr-Newman-Kasuya_case2_x}
\end{equation}
\begin{equation}
\Psi_{out}(r>r_{+})=\mbox{e}^{-i \omega t}(r-r_{+})^{\frac{i}{2\kappa_{+}}(\omega-\omega_{0})}\ .
\label{eq:sol_out_2_Kerr-Newman-Kasuya_case2_x}
\end{equation}
The relative probability of creating a particle-antiparticle pair just outside the horizon is given by
\begin{equation}
\Gamma_{+}=\left|\frac{\Psi_{out}(r>r_{+})}{\Psi_{out}(r<r_{+})}\right|^{2}=\mbox{e}^{-\frac{2\pi}{\kappa_{+}}(\omega-\omega_{0})}\ ,
\label{eq:taxa_refl_Kerr-Newman-Kasuya_case2_x}
\end{equation}
and the resulting Hawking radiation spectrum for scalar particles being radiated from a Kerr-Newman-Kasuya black hole is given by \cite{AnnPhys.362.576}
\begin{equation}
\left|N_{\omega}\right|^{2}=\frac{\Gamma_{+}}{1-\Gamma_{+}}=\frac{1}{\mbox{e}^{\frac{\hbar(\omega-\omega_{0})}{k_{B}T_{+}}}-1}\ ,
\label{eq:espectro_rad_Kerr-Newman-Kasuya_case2_x_2_2}
\end{equation}
where $\omega_{0}=m\Omega_{+}+e\Phi_{+}$.
%
%%%%%%%%%%%%%%%%%%%%%%%%%%%%%%%%%%%%%%%%%%%%%%%%%%%%%%%%%%%%%%%%%%%%%%%%%%%%%%%%%%%%%%%%%%%%%% Special cases
%
\subsection{Special cases}
In the last section we have used the conventional forms of the series expansions and the asymptotic forms of the confluent Heun functions, $\mbox{HeunC}(\alpha$, $\beta$, $\gamma$, $\delta$, $\eta$; $x$), to obtain the radial solution.

Now, from this general exact solution given by Eq.~(\ref{eq:sol_scattering_wave_Kerr-Newman-Kasuya_case2_x_2}), we can write down some particular results for the Kerr and Schwarzschild spacetimes.
%
%%%%%%%%%%%%%%%%%%%%%%%%%%%%%%%%%%%%%%%%%%%%%%%%%%%%%%%%%%%%%%%%%%%%%%%%%%%%%%%%%%%%%%%%%%%%%% The Kerr black hole
%
\subsubsection{The Kerr black hole}
In the special case of $Q_{e}=Q_{m}=0$ the metric given by Eq.~(\ref{eq:metrica_Kerr-Newman-Kasuya}) reduces to the Kerr form. Accordingly, we have
\begin{equation}
\Delta=r^{2}-2Mr+a^{2}\ ,
\label{eq:Delta_metrica_Kerr}
\end{equation}
\begin{equation}
r_{\pm}=M \pm (M^{2}-a^{2})^{1/2}\ .
\label{eq:sol_padrao_Kerr}
\end{equation}

Then, from Eqs.~(\ref{eq:alpha_radial_Kerr-Newman-Kasuya_case2_x}) and (\ref{eq:delta_radial_Kerr-Newman-Kasuya_case2_x}), we get
\begin{equation}
-i\frac{1}{2}\alpha=k_{0} (r_{+}-r_{-})\ ,
\label{eq:-ialpha/2_radial_Kerr_case2_x}
\end{equation}
\begin{equation}
i\frac{\delta}{\alpha}=\gamma_{0}\ ,
\label{eq:idelta/alpha_radial_Kerr_case2_x}
\end{equation}
where $k_{0}^{2} \equiv \omega ^2-\mu_{0} ^2$ and $\gamma_{0} \equiv (M\mu_{0} ^2-2M \omega ^2)/(\omega ^2-\mu_{0} ^2)^{\frac{1}{2}}$.

Therefore, the analytical solution for the regular partial wave, in this case, is given by
\begin{equation}
R_{l}(r) \sim C_{l}\ \frac{1}{r} \sin[k_{0}r-\gamma_{0}\ln r+\sigma_{l}(\omega)]\ ,
\label{eq:sol_scattering_wave_Kerr_case2_x_2}
\end{equation}
where the eigenvalues $\lambda=\lambda_{lm}$ ($l,m$ are integers such that $|m| \leq l$) are not expressible in analytic form in terms of $l$ and $m$.
%
%%%%%%%%%%%%%%%%%%%%%%%%%%%%%%%%%%%%%%%%%%%%%%%%%%%%%%%%%%%%%%%%%%%%%%%%%%%%%%%%%%%%%%%%%%%%%% Massless scalar fields
%
%\subsubsection{Massless scalar fields}

If we particularize this result for the massless case, that is, $\mu_{0}=0$, we have
\begin{equation}
k_{0}=\omega\ ,
\label{eq:-ialpha/2_radial_Kerr_case2_x_massless}
\end{equation}
\begin{equation}
\gamma_{0}=-2M\omega\ .
\label{eq:idelta/alpha_radial_Kerr_case2_x_massless}
\end{equation}

Now, using the definition of the tortoise coordinate, $r_{*}$, namely,
\begin{equation}
\frac{dr}{dr_{*}}=\frac{\Delta}{r^{2}+a^{2}}\ ,
\label{eq:coord_tortoise_1}
\end{equation}
when $r \rightarrow \infty$ we get
\begin{equation}
r_{*}=r+2M\ln r\ .
\label{eq:coord_tortoise_2_infty}
\end{equation}

In this new coordinate, we obtain the following scattering wave solution for massless scalar particles in the Kerr spacetime
\begin{equation}
R_{\lambda}(r) \sim \frac{1}{r} (A\ \mbox{e}^{i \omega r_{*}}+B\ \mbox{e}^{-i \omega r_{*}})\ ,
\label{eq:sol_scattering_wave_Kerr_case2_x_2_massless}
\end{equation}
where $A$ and $B$ are constants and describe the incident and reflected wave coefficients, respectively. The solution (\ref{eq:sol_scattering_wave_Kerr_case2_x_2_massless}) is exactly the solution obtained by Starobinsky \cite{ZhEkspTeorFiz.64.48}, Handler and Matzner \cite{PhysRevD.22.2331}, and Glampedakis and Andersson \cite{ClassQuantumGrav.18.1939}.

Furthermore, from Eq.~(\ref{eq:quasispectrum_modes_Kerr-Newman-Kasuya_case2_x}) we can write the expression of the resonant frequencies for massless scalar fields in the Kerr spacetime, which is given by
\begin{equation}
\omega_{R}=\frac{\omega_{+}\kappa_{-}+\omega_{-}\kappa_{+}}{4M\kappa_{+}\kappa_{-}+\kappa_{+}+\kappa_{-}}\ ,
\label{eq:omegaR_Kerr_case2_x}
\end{equation}
\begin{equation}
\omega_{I}=\frac{2\kappa_{+}\kappa_{-}(n+1)}{4M\kappa_{+}\kappa_{-}+\kappa_{+}+\kappa_{-}}\ ,
\label{eq:omegaI_Kerr_case2_x}
\end{equation}
where $n \geq 0$. The solutions (\ref{eq:omegaR_Kerr_case2_x}) and (\ref{eq:omegaI_Kerr_case2_x}) generalize the one obtained by Felice \cite{PhysRevD.19.451} in the sense that we have an analytic solution valid over the range $0 < \omega < \infty$ and $|m| \leq l$, contrary to the result presented in the literature \cite{PhysRevD.19.451} which was obtained for high angular momentum and low energy (frequency).
%
%%%%%%%%%%%%%%%%%%%%%%%%%%%%%%%%%%%%%%%%%%%%%%%%%%%%%%%%%%%%%%%%%%%%%%%%%%%%%%%%%%%%%%%%%%%%%% The Schwarzschild black hole
%
\subsubsection{The Schwarzschild black hole}
In this case, the metric (\ref{eq:metrica_Kerr-Newman-Kasuya}) reduces to the Schwarzschild form when we take $a=Q_{e}=Q_{m}=0$:
\begin{equation}
\Delta=r^{2}-2Mr\ ,
\label{eq:Delta_metrica_Schwarzschild}
\end{equation}
\begin{equation}
r_{h}=2M\ .
\label{eq:sol_padrao_Schwarzschild}
\end{equation}

Thus, we have
\begin{equation}
k_{0}^{2}=\omega ^2-\mu_{0} ^2\ ,
\label{eq:-ialpha/2_radial_Schwarzschild_case2_x}
\end{equation}
\begin{equation}
\gamma_{0}=\frac{M\mu_{0} ^2-2M \omega ^2}{(\omega ^2-\mu_{0} ^2)^{\frac{1}{2}}}\ .
\label{eq:idelta/alpha_radial_Schwarzschild_case2_x}
\end{equation}

Therefore, the analytical solution for the regular partial wave has the asymptotic form
\begin{equation}
R_{l}(r) \sim C_{l}\ \frac{1}{r} \sin\left[k_{0}r-\gamma_{0}\ln r-\frac{l\pi}{2}+\sigma_{l}(\omega)\right]\ ,
\label{eq:sol_scattering_wave_Schwarzschild_case2_x_2}
\end{equation}
where $\lambda=l(l+1)$, and the phase shift $\sigma_{l}$ is given by
\begin{equation}
\sigma_{l}(\omega)=\arg\Gamma(l+1+i\gamma_{0})\ .
\label{eq:phase_shift_scattering_wave_Schwarzschild_case2_x_2}
\end{equation}
The solution (\ref{eq:sol_scattering_wave_Schwarzschild_case2_x_2}) is exactly the solution obtained by Kofinti \cite{IntJTheorPhys.23.991}.
%
%%%%%%%%%%%%%%%%%%%%%%%%%%%%%%%%%%%%%%%%%%%%%%%%%%%%%%%%%%%%%%%%%%%%%%%%%%%%%%%%%%%%%%%%%%%%%% Massless scalar fields
%
%\subsubsection{Massless scalar fields}

Once more, considering a massless scalar field, we have
\begin{equation}
k_{0}=\omega\ ,
\label{eq:-ialpha/2_radial_Schwarzschild_case2_x_massless}
\end{equation}
\begin{equation}
\gamma_{0}=-2M\omega\ ,
\label{eq:idelta/alpha_radial_Schwarzschild_case2_x_massless}
\end{equation}
\begin{equation}
R_{l}(r) \sim C_{l}\ \frac{1}{r} \sin\left[\omega r+2M\omega\ln r-\frac{l\pi}{2}+\delta_{l}(\omega)\right]\ .
\label{eq:sol_scattering_wave_Schwarzschild_case2_x_2_massless}
\end{equation}
This is exactly the solution obtained by Sanchez \cite{JMathPhys.17.688}, where the phase shift $\delta_{l}(\omega)$ was obtained by JWKB approximation.

Finally, from Eq.~(\ref{eq:quasispectrum_modes_Kerr-Newman-Kasuya_case2_x}), the resonant frequencies for massless scalar fields in the Schwarzschild spacetime are given by
\begin{equation}
\omega_{QS}=\frac{i}{4M}(n+1)\ ,
\label{eq:quasispectrum_modes_Schwarzschild_case2_x}
\end{equation}
with $n \geq 0$.
%
%%%%%%%%%%%%%%%%%%%%%%%%%%%%%%%%%%%%%%%%%%%%%%%%%%%%%%%%%%%%%%%%%%%%%%%%%%%%%%%%%%%%%%%%%%%%%% Massless scalar field equation in the Ernst spacetime
%
\section{Massless scalar field equation in the Ernst spacetime}
The spacetime generated by a black hole with electric charge $Q$ surrounded by a magnetic field, which corresponds effectively to the Reissner-Nordstr\"{o}m black hole surrounded by a magnetic field, is given by the Ernst metric \cite{JMathPhys.17.54}, whose line element takes the following form
\begin{equation}
ds^{2}=|\Lambda|^2\left(-\frac{\Delta}{r^{2}}\ dt^{2}+\frac{r^2}{\Delta}\ dr^{2}+r^{2}\ d\theta^{2}\right)+\frac{r^{2}\sin^2\theta}{|\Lambda|^2}(d\phi-\omega'\ dt)^{2}\ ,
\label{eq:metrica_Ernst}
\end{equation}
with
\begin{equation}
\Delta=r^{2}-2Mr+Q^{2}\ ,
\label{eq:Delta_metrica_Ernst}
\end{equation}
\begin{equation}
\Lambda=1+\frac{1}{4}B_{0}^{2}(r^{2}\sin^{2}\theta+Q^{2}\cos^{2}\theta)-iB_{0}Q\cos\theta\ ,
\label{eq:Lambda_metrica_Ernst}
\end{equation}
\begin{equation}
\omega'=-\frac{2B_{0}Q}{r}+B_{0}^{3}\left[Qr+\frac{1}{2}\frac{Q^{3}}{r}-\frac{1}{2}\frac{Q}{r}(r^{2}-2Mr+Q^{2})\sin^{2}\theta\right]\ ,
\label{eq:omega_metrica_Ernst}
\end{equation}
where the external magnetic field is determined by the parameter $B_{0}$.

The horizon surface equation is obtained from the condition $\Delta=(r-r_{+})(r-r_{-})=0$, whose solutions correspond to the event (and Cauchy) horizons and are given by
\begin{equation}
r_{\pm}=M\pm(M^{2}-Q^{2})^{\frac{1}{2}}\ .
\label{eq:sol_Delta_metrica_Ernst}
\end{equation}
Furthermore, the surface area of the horizon in the presence of a external magnetic field, $\mathcal{A}_{+,B_{0}}$, is given by
\begin{eqnarray}
\mathcal{A}_{+,B_{0}} & = & \left.\int\int\sqrt{-g}\ d\theta\ d\phi \right|_{r=r_{+}}\nonumber\\
& = & 4 \pi r_{+}^{2}\left[1+B_{0}^{2}\left(\frac{Q}{2}+\frac{r_{+}^{2}}{3}\right)\right]+\mathcal{O}(B_{0}^{4})\ ,
\label{eq:area_Ernst}
\end{eqnarray}
where $g \equiv \mbox{det}(g_{\sigma\tau})=-r^{4}|\Lambda|^{4}\sin^{2}\theta$.

In order to solve the Klein-Gordon equation for a massless field without interacting with an electromagnetic field in the background under consideration, let us first estimate a real astrophysical situation. This corresponds to match the Ernst solution with a Schwarzschild solution at some large $r$, which is done by neglecting terms higher than $B_{0}^{2}$. These terms are growing far from black hole, which means that in the asymptotically far region the gravitational field is divergent. Thus, the Klein-Gordon equation can be written in the spacetime (\ref{eq:metrica_Ernst}) as
\begin{eqnarray}
& & \biggl\{\frac{4B_{0}Qr^{3}}{\Delta}\frac{\partial^{2}}{\partial t\ \partial\phi}-\frac{r^{4}}{\Delta}\frac{\partial^{2}}{\partial t^{2}}+\biggl[\frac{1}{\sin ^{2}\theta}-\frac{4B_{0}^{2}Q^{2}r^{2}}{\Delta}\nonumber\\
& & +B_{0}^{2}(3Q^{2}\cot^{2}\theta+r^{2})\biggl]\frac{\partial^{2}}{\partial\phi^{2}}+\frac{\partial}{\partial r}\left(\Delta\frac{\partial}{\partial r}\right)+\frac{1}{\sin\theta}\frac{\partial}{\partial\theta}\left(\sin\theta\frac{\partial}{\partial\theta}\right)\biggr\}\Psi=0\ .\nonumber\\
\label{eq:mov_Ernst}
\end{eqnarray}

We consider, once again, for the same reasons considered previously, the solution of Eq.~(\ref{eq:mov_Ernst}) as having the following form
\begin{equation}
\Psi=\Psi(\mathbf{r},t)=R(r)S(\theta)\mbox{e}^{im\phi}\mbox{e}^{-i\omega t}\ ,
\label{eq:separacao_variaveis_Ernst}
\end{equation}
where $m=\pm 1,\pm 2,\pm 3,...$ is the angular momentum quantum number, and the frequency is taken positive, that is, $\omega > 0$. Substituting Eq.~(\ref{eq:separacao_variaveis_Ernst}) into (\ref{eq:mov_Ernst}), we find that the functions $S(\theta)$ and $R(r)$ satisfy the following equations
\begin{equation}
\frac{1}{\sin\theta}\frac{d}{d\theta}\left(\sin\theta\frac{dS}{d\theta}\right)+\biggl(\lambda-\frac{m^{2}}{\sin^{2}\theta}-3B_{0}^{2}Q^{2}m^{2}\cot^{2}\theta\biggr)S=0
\label{eq:mov_angular_Ernst}
\end{equation}
and
\begin{equation}
\frac{d}{dr}\left(\Delta\frac{dR}{dr}\right)+\biggl[\frac{1}{\Delta}(2B_{0}Qmr+\omega r^{2})^{2}-(\lambda+B_{0}^{2}m^{2}r^{2})\biggr]R=0\ ,
\label{eq:mov_radial_Ernst}
\end{equation}
where $\lambda$ is the separation constant.
%
%%%%%%%%%%%%%%%%%%%%%%%%%%%%%%%%%%%%%%%%%%%%%%%%%%%%%%%%%%%%%%%%%%%%%%%%%%%%%%%%%%%%%%%%%%%%%% The angular equation
%
\subsection{The angular equation}
In order to obtain the analytical solution of the angular Klein-Gordon equation, let us perform a change of variable such that
\begin{equation}
x=\cos^{2}\theta\ .
\label{eq:coord_angular_Ernst_x}
\end{equation}
With this transformation, Eq.~(\ref{eq:mov_angular_Ernst}) can be written as
\begin{equation}
\frac{d^{2}S(x)}{dx^{2}}+\left(\frac{1/2}{x}+\frac{1}{x-1}\right)\frac{dS(x)}{dx}+\biggl[\frac{A_{1}}{x}+\frac{A_{2}}{x-1}-\frac{(A_{3})^{2}}{(x-1)^2}\biggr]S(x)=0\ ,
\label{eq:mov_angular_Ernst_x}
\end{equation}
where the coefficients $A_{1}$, $A_{2}$, and $A_{3}$ are given by
\begin{equation}
A_{1}=\frac{m^2-\lambda}{4}\ ,
\label{eq:A1_angular_Ernst_x}
\end{equation}
\begin{equation}
A_{2}=\frac{\lambda-m^2}{4}\ ,
\label{eq:A2_angular_Ernst_x}
\end{equation}
\begin{equation}
A_{3}=i\frac{\left(3 B_{0}^2 m^2 Q^2+m^2\right)^{\frac{1}{2}}}{2}\ .
\label{eq:A3_angular_Ernst_x}
\end{equation}

This equation for the dependent variable $S(x)$ is converted into a Heun-type equation for $U$ by a \textit{F-homotopic transformation} of the type
\begin{equation}
S(x)=(x-1)^{A_{3}}U(x)\ .
\label{eq:F-homotopic_mov_angular_Ernst_x}
\end{equation}
Explicitly, the result of applying (\ref{eq:F-homotopic_mov_angular_Ernst_x}) to the angular equation written as (\ref{eq:mov_angular_Ernst_x}) is
\begin{eqnarray}
& & \frac{d^{2}U(x)}{dx^{2}}+\left(\frac{1/2}{x}+\frac{2 A_{3}+1}{x-1}\right)\frac{dU(x)}{dx}\nonumber\\
& & +\left(\frac{A_{1}-A_{3}/2}{x}+\frac{A_{2}+A_{3}/2}{x-1}\right)U(x)=0\ .
\label{eq:mov_angular_Ernst_x_U}
\end{eqnarray}

Thus, the linearly independent general exact solution of the angular Klein-Gordon equation for a massless scalar field in the Reissner-Nordstr\"{o}m black hole surrounded by a magnetic field (Ernst spacetime), over the entire range $0 \leq x < \infty$, can be written as
\begin{eqnarray}
S(x) & = & (x-1)^{\frac{1}{2}\gamma}\{C_{1}\ \mbox{HeunC}(\alpha,\beta,\gamma,\delta,\eta;x)\nonumber\\
& & +C_{2}\ x^{-\beta}\ \mbox{HeunC}(\alpha,-\beta,\gamma,\delta,\eta;x)\}\ ,
\label{eq:solucao_geral_angular_Ernst_x}
\end{eqnarray}
where $C_{1}$ and $C_{2}$ are constants, and the parameters $\alpha$, $\beta$, $\gamma$, $\delta$, and $\eta$ are given by
\begin{equation}
\alpha=0\ ,
\label{eq:alpha_angular_Ernst_x}
\end{equation}
\begin{equation}
\beta=-\frac{1}{2}\ ,
\label{eq:beta_angular_Ernst_x}
\end{equation}
\begin{equation}
\gamma=im(3 B_{0}^2 Q^2+1)^{\frac{1}{2}}\ ,
\label{eq:gamma_angular_Ernst_x}
\end{equation}
\begin{equation}
\delta=0\ ,
\label{eq:delta_angular_Ernst_x}
\end{equation}
\begin{equation}
\eta=\frac{1}{4}(1-m^{2}+\lambda)\ .
\label{eq:eta_angular_Ernst_x}
\end{equation}
%
%%%%%%%%%%%%%%%%%%%%%%%%%%%%%%%%%%%%%%%%%%%%%%%%%%%%%%%%%%%%%%%%%%%%%%%%%%%%%%%%%%%%%%%%%%%%%% The radial equation
%
\subsection{The radial equation}
Now, in order to obtain the analytical solution of the radial Klein-Gordon equation, let us use Eq.~(\ref{eq:sol_Delta_metrica_Ernst}) to write down Eq.~(\ref{eq:mov_radial_Ernst}) as
\begin{eqnarray}
&& \frac{d^{2}R(r)}{dr^{2}}+\left(\frac{1}{r-r_{+}}+\frac{1}{r-r_{-}}\right)\frac{dR(r)}{dr}+\frac{1}{(r-r_{+})(r-r_{-})}\biggl\{\nonumber\\
&& r(4 B_{0} m Q+r_{+} \omega +r_{-} \omega )\omega-r^2 (B_{0}^2 m^2-\omega ^2)+4\biggl[B_{0}mQ-\frac{\omega}{2}(r_{+}+r_{-})\biggr]^{2}\nonumber\\
&& -\omega^{2}r_{+}r_{-}-\lambda+\frac{ (2 B_{0} m Qr_{+}+\omega r_{+}^2)^2}{(r-r_{+}) (r_{+}-r_{-})}+\frac{(2 B_{0} m Qr_{-}+\omega r_{-}^2)^2}{(r-r_{-}) (r_{-}-r_{+})}\biggr\}R(r)=0\ .\nonumber\\
\label{eq:mov_radial_Ernst_r}
\end{eqnarray}
Since this equation has singularities at $r=(r_{+},r_{-},\infty)$, by the homographic substitution
\begin{equation}
x=\frac{r-r_{+}}{r_{-}-r_{+}}\ ,
\label{eq:homog_subs_radial_Ernst_x}
\end{equation}
we bring Eq.~(\ref{eq:mov_radial_Ernst_r}) into the Heun's equation form given by
\begin{eqnarray}
&& \frac{d^{2}R(x)}{dx^{2}}+\left(\frac{1}{x}+\frac{1}{x-1}\right)\frac{dR(x)}{dx}\nonumber\\
&& +\biggl[-(D_{1})^{2} +\frac{D_{2}}{x}+\frac{D_{3}}{x-1}-\frac{(D_{4})^{2}}{x^2}-\frac{(D_{5})^{2}}{(x-1)^2}\biggr]R(x)=0\ ,
\label{eq:mov_radial_Ernst_x}
\end{eqnarray}
where the coefficients $D_{1}$, $D_{2}$, $D_{3}$, $D_{4}$, and $D_{5}$ are expressed as
\begin{equation}
D_{1}=i(r_{+}-r_{-})(\omega ^2-B_{0}^2 m^2)^{\frac{1}{2}}\ ,
\label{eq:D1_radial_Ernst_x}
\end{equation}
\begin{eqnarray}
D_{2} & = & -\frac{2 r_{+}^2 \omega  (2 B_{0} m Q r_{+}-6 B_{0} m Q r_{-}+r_{+}^2 \omega)}{(r_{+}-r_{-})^2 }\nonumber\\
&& +\frac{B_{0}^2 m^2 r_{+} (8 Q^2 r_{-}+r_{+}^3-2 r_{+}^2 r_{-}+r_{+} r_{-}^2)}{(r_{+}-r_{-})^2}\nonumber\\
&& +\frac{4 r_{+}^3r_{-} \omega^2+\lambda  (r_{+}-r_{-})^2}{(r_{+}-r_{-})^2 }\ ,
\label{eq:D2_radial_Ernst_x}
\end{eqnarray}
\begin{eqnarray}
D_{3} & = & -\frac{2 r_{-}^2 \omega  (6 B_{0} m Q r_{+}-2 B_{0} m Q r_{-}-r_{-}^2\omega)}{(r_{+}-r_{-})^2 }\nonumber\\
&& -\frac{B_{0}^2 m^2 r_{-} (8 Q^2 r_{+}+r_{+}^2 r_{-}-2 r_{+} r_{-}^2+r_{-}^3)}{(r_{+}-r_{-})^2 }\nonumber\\
&& -\frac{4 r_{-}^3r_{+} \omega^2+\lambda  (r_{+}-r_{-})^2}{(r_{+}-r_{-})^2 }\ ,
\label{eq:D3_radial_Ernst_x}
\end{eqnarray}
\begin{equation}
D_{4}=i\frac{(2 B_{0} m Qr_{+}+\omega r_{+}^{2})}{ r_{+}-r_{-}}\ ,
\label{eq:D4_radial_Ernst_x}
\end{equation}
\begin{equation}
D_{5}=i\frac{(2 B_{0} m Qr_{-}+\omega r_{-}^{2})}{ r_{+}-r_{-}}\ .
\label{eq:D5_radial_Ernst_x}
\end{equation}

Having thus moved the singularities to the points $x=0,1$, now we apply the specialized form of the \textit{s-homotopic transformation} of the dependent variable $R(x) \mapsto U(x)$, namely
\begin{equation}
R(x)=\mbox{e}^{D_{1}x}x^{D_{4}}(x-1)^{D_{5}}U(x)\ .
\label{eq:s-homotopic_mov_radial_Ernst_x}
\end{equation}
Using this transformation, Eq.~(\ref{eq:mov_radial_Ernst_x}) turns into an equation for $U(x)$, which can be written as
\begin{eqnarray}
&& \frac{d^{2}U(x)}{dx^{2}}+\left(2 D_{1}+\frac{2 D_{4}+1}{x}+\frac{2 D_{5}+1}{x-1}\right)\frac{dU(x)}{dx}\nonumber\\
&& +\biggl[\frac{D_{2}( r_{-}-r_{+})+(2 D_{4}+1)(- D_{5}+ D_{1})-D_{4}}{x}\nonumber\\
&& +\frac{D_{3}( r_{-}-r_{+})+(2 D_{5}+1)(D_{4}+ D_{1})}{x-1}+\frac{D_{5}}{x-1}\biggr]U(x)=0\ .
\label{eq:mov_radial_Ernst_x_U}
\end{eqnarray}

Thus, the linearly independent general exact solution of the radial Klein-Gordon equation for a massless scalar field in the Reissner-Nordstr\"{o}m black hole surrounded by a magnetic field, over the entire range $0 \leq x < \infty$, can be written as
\begin{eqnarray}
R(x) & = & \mbox{e}^{\frac{1}{2}\alpha x}x^{\frac{1}{2}\beta}(x-1)^{\frac{1}{2}\gamma}\{C_{1}\ \mbox{HeunC}(\alpha,\beta,\gamma,\delta,\eta;x)\nonumber\\
&& +C_{2}\ x^{-\beta}\ \mbox{HeunC}(\alpha,-\beta,\gamma,\delta,\eta;x)\}\ ,
\label{eq:solucao_geral_radial_Ernst_x}
\end{eqnarray}
where $C_{1}$ and $C_{2}$ are constants, and the parameters $\alpha$, $\beta$, $\gamma$, $\delta$, and $\eta$ are given by
\begin{equation}
\alpha=2i(r_{+}-r_{-})(\omega ^2-B_{0}^2 m^2)^{\frac{1}{2}}\ ,
\label{eq:alpha_radial_Ernst_x}
\end{equation}
\begin{equation}
\beta=2i\frac{(2 B_{0} m Qr_{+}+\omega r_{+}^{2})}{ r_{+}-r_{-}}\ ,
\label{eq:beta_radial_Ernst_x}
\end{equation}
\begin{equation}
\gamma=2i\frac{(2 B_{0} m Qr_{-}+\omega r_{-}^{2})}{ r_{+}-r_{-}}\ ,
\label{eq:gamma_radial_Ernst_x}
\end{equation}
\begin{equation}
\delta=(r_{+}-r_{-})(r_{+}+r_{-})(B_{0}^2 m^2-4 B_{0} m Q \omega -2 \omega ^2)\ ,
\label{eq:delta_radial_Ernst_x}
\end{equation}
\begin{eqnarray}
\eta & = & \frac{2 r_{+}^2 \omega  \left(2 B_{0} m Q r_{+}-6 B_{0} m Q r_{-}+r_{+}^2 \omega\right)}{(r_{+}-r_{-})^2 }\nonumber\\
&& -\frac{B_{0}^2 m^2 r_{+} \left(8 Q^2 r_{-}+r_{+}^3-2 r_{+}^2 r_{-}+r_{+} r_{-}^2\right)}{(r_{+}-r_{-})^2 }\nonumber\\
&& -\frac{4 r_{+}^3r_{-} \omega^2 +\lambda  (r_{+}-r_{-})^2}{(r_{+}-r_{-})^2 }\ .
\label{eq:eta_radial_Ernst_x}
\end{eqnarray}

Note the dependence of both angular and radial solutions with the parameter $B_{0}$, associated with the external magnetic field.
%
%%%%%%%%%%%%%%%%%%%%%%%%%%%%%%%%%%%%%%%%%%%%%%%%%%%%%%%%%%%%%%%%%%%%%%%%%%%%%%%%%%%%%%%%%%%%%% Resonant frequencies
%
\subsection{Resonant frequencies}
Imposing the so called $\delta$-condition given by Eq.~(\ref{eq:cond_polin_1}) to the radial part $R(r)$ of Eq.~(\ref{eq:separacao_variaveis_Ernst}), we obtain the following expression for the resonant frequencies
\begin{eqnarray}
&& \frac{M(B_{0}^{2}m^{2}-4B_{0}mQ\omega-2\omega^{2})}{(B_{0}^{2}m^{2}-\omega^{2})^{\frac{1}{2}}}\nonumber\\
&& +\frac{i}{2}\left[\left(\frac{\omega}{\kappa_{+}}+\frac{\omega}{\kappa_{-}}\right)+\left(\frac{\omega_{+,B_{0}}}{\kappa_{+}}+\frac{\omega_{-,B_{0}}}{\kappa_{-}}\right)\right]=-(n+1)\ ,
\label{eq:quasinormal_modes_Ernst_x}
\end{eqnarray}
where 
\begin{equation}
\kappa_{\pm} \equiv \frac{1}{2}\frac{1}{r_{\pm}^{2}}\left.\frac{d\Delta}{dr}\right|_{r=r_{\pm}}=\frac{1}{2}\frac{r_{\pm}-r_{\mp}}{r_{\pm}^{2}}\ ,
\label{eq:acel_grav_ext_Ernst}
\end{equation}
\begin{equation}
\omega_{\pm,B_{0}}=m\tilde{\Omega}_{\pm,B_{0}}\ ,
\label{eq:omega_Ernst}
\end{equation}
\begin{equation}
\tilde{\Omega}_{\pm,B_{0}}=\left.-\frac{g_{03}}{g_{33}}\right|_{r=r_{\pm}}=\frac{2B_{0}Q}{r_{\pm}}+\mathcal{O}(B_{0}^{3})\ .
\label{eq:vel_ang_Ernst}
\end{equation}

It is not possible to obtain an analytic expression for $\omega_{n}$ from Eq.~(\ref{eq:quasinormal_modes_Ernst_x}), however, there are several numerical methods that can be used to obtain approximate expressions for each energy level \cite{PhysRevD.84.127502}.
%
%%%%%%%%%%%%%%%%%%%%%%%%%%%%%%%%%%%%%%%%%%%%%%%%%%%%%%%%%%%%%%%%%%%%%%%%%%%%%%%%%%%%%%%%%%%%%% Scattering
%
\subsection{Scattering}
If we consider the asymptotic series given by Eq.~(\ref{eq:assintotica_HeunC_x_grande}), the radial wave function can be written as
\begin{equation}
R_{\lambda}(x) \sim C_{\lambda}\ \frac{1}{x} \sin\left[-i\frac{1}{2}\alpha x+i\frac{\delta}{\alpha}\ln x+\sigma_{\lambda}(\omega)\right]\ ,
\label{eq:scattering_wave_Ernst_x_2}
\end{equation}
where $\sigma_{\lambda}(\omega)$ is the phase shift.

From Eqs.~(\ref{eq:alpha_radial_Ernst_x}) and (\ref{eq:delta_radial_Ernst_x}), we obtain
\begin{equation}
-i\frac{\alpha}{2}=k_{0,B_{0}} (r_{+}-r_{-})\ ,
\label{eq:-ialpha/2_radial_Ernst_x}
\end{equation}
\begin{equation}
i\frac{\delta}{\alpha}=\gamma_{0,B_{0}}\ ,
\label{eq:idelta/alpha_radial_Ernst_x}
\end{equation}
where
\begin{equation}
k_{0,B_{0}}^{2} \equiv \omega ^2-B_{0}^{2}m^{2}\ ,
\label{eq:k0_radial_Ernst_x}
\end{equation}
\begin{equation}
\gamma_{0,B_{0}} \equiv \frac{MB_{0}^{2}m^{2}-4MB_{0}mQ\omega-2M \omega ^2}{(\omega ^2-B_{0}^{2}m^{2})^{\frac{1}{2}}}\ .
\label{eq:gamma0_radial_Ernst_x}
\end{equation}
Therefore, the regular partial wave solution has the asymptotic form
\begin{equation}
R_{\lambda}(r) \sim C_{\lambda}\ \frac{1}{r} \sin[k_{0,B_{0}}r-\gamma_{0,B_{0}}\ln r+\sigma_{\lambda}(\omega)]\ .
\label{eq:sol_scattering_wave_Ernst_x_2}
\end{equation}

These solutions for the scalar fields far from the black hole can be useful to investigate the scattering of massless scalar particles. It is worth calling attention to the fact that we are using the analytical solution of the radial part of the Klein-Gordon equation in the spacetime under consideration.

Indeed, the phase shift, $\sigma_{\lambda}$, is not a simple function of $\lambda$, and thus, the exact expression of the scattering amplitude is not available but just an approximate one is obtained, as is shown for instance by Abramov et al. \cite{JPhysBAtomMolecPhys.12.1761}.
%
%%%%%%%%%%%%%%%%%%%%%%%%%%%%%%%%%%%%%%%%%%%%%%%%%%%%%%%%%%%%%%%%%%%%%%%%%%%%%%%%%%%%%%%%%%%%%% Hawking radiation
%
\subsection{Hawking radiation}
The radial solution given by Eq.~(\ref{eq:solucao_geral_radial_Ernst_x}) has the following asymptotic behavior at the exterior event horizon $r_{+}$ \cite{AnnPhys.350.14}:
\begin{equation}
R(r) \sim C_{1}\ (r-r_{+})^{\frac{\beta}{2}}+C_{2}\ (r-r_{+})^{-\frac{\beta}{2}}\ ,
\label{eq:exp_0_solucao_geral_radial_Ernst_x}
\end{equation}
where all constants involved are included in $C_{1}$ and $C_{2}$.

From Eq.~(\ref{eq:beta_radial_Ernst_x}), the parameter $\beta$ can be written as
\begin{equation}
\frac{\beta}{2}=\frac{i}{2\kappa_{+}}(\omega+\omega_{0,B_{0}})=\frac{i}{2\kappa_{+}}\tilde{\omega}\ ,
\label{eq:expoente_rad_Hawking_Ernst}
\end{equation}
where $\omega_{0,B_{0}}=m\tilde{\Omega}_{+,B_{0}}$, being the surface gravity on the background horizon surface, $\kappa_{+}$, and the `dragging angular velocity of the exterior horizon', $\tilde{\Omega}_{+,B_{0}}$, given by Eqs.~(\ref{eq:acel_grav_ext_Ernst}) and (\ref{eq:vel_ang_Ernst}), respectively.

Therefore, considering the time factor, on the black hole exterior horizon surface, the ingoing and outgoing wave solutions are
\begin{equation}
\Psi_{in}=\mbox{e}^{-i \omega t}(r-r_{+})^{-\frac{i}{2\kappa_{+}}\tilde{\omega}}\ ,
\label{eq:sol_in_1_Ernst}
\end{equation}
\begin{equation}
\Psi_{out}(r>r_{+})=\mbox{e}^{-i \omega t}(r-r_{+})^{\frac{i}{2\kappa_{+}}\tilde{\omega}}\ .
\label{eq:sol_out_2_Ernst}
\end{equation}
These solutions depend on the parameter $B_{0}$, in such a way that the total energy of the radiated particles is increased due to the presence of the external magnetic field.

Following the same procedure developed in a recent paper \cite{EurophysLett.109.60006}, the relative scattering probability of the scalar wave at the event horizon surface, $\Gamma_{+}$, and the Hawking radiation spectrum of scalar particles, $\left|N_{\omega}\right|^{2}$, respectively, are given by
\begin{equation}
\Gamma_{+}=\left|\frac{\Psi_{out}(r>r_{+})}{\Psi_{out}(r<r_{+})}\right|^{2}=\mbox{e}^{-\frac{2\pi}{\kappa_{+}}\tilde{\omega}}\ ,
\label{eq:taxa_refl_Ernst}
\end{equation}
\begin{equation}
\left|N_{\omega}\right|^{2}=\frac{1}{\mbox{e}^{\frac{2\pi}{\kappa_{+}}\tilde{\omega}}-1}=\frac{1}{\mbox{e}^{\frac{\hbar}{k_{B}T_{+}}(\omega+\omega_{0,B_{0}})}-1}\ ,
\label{eq:espectro_rad_Ernst_2}
\end{equation}
with $T_{+}=\hbar\kappa_{+}/2\pi k_{B}$ being the Hawking radiation temperature.

Therefore, we can see that the resulting Hawking radiation spectrum of scalar particles has a thermal character. It is worth noticing that the total energy of radiated scalar particles is increased due to the presence of the external magnetic field, more precisely, this gives rise to a kind of `dragging angular velocity of the exterior horizon', $\tilde{\Omega}_{+,B_{0}}$, in comparison with the scenario without an external magnetic field \cite{AnnPhys.362.576}.
%
%%%%%%%%%%%%%%%%%%%%%%%%%%%%%%%%%%%%%%%%%%%%%%%%%%%%%%%%%%%%%%%%%%%%%%%%%%%%%%%%%%%%%%%%%%%%%% Special case: Schwarzschild black hole with a magnetic field
%
\subsection{Special case: Schwarzschild black hole with a magnetic field}
We now examine the special case of an uncharged black hole $Q=0$. Thus, the metric (\ref{eq:metrica_Ernst}) reduces to the Schwarzschild form with a magnetic field. Accordingly, we have
\begin{equation}
\Delta=r^{2}-2Mr\ ,
\label{eq:Delta_metrica_Ernst_Schwarzschild}
\end{equation}
\begin{equation}
\Lambda=1+\frac{1}{4}B_{0}^{2}r^{2}\sin^{2}\theta\ ,
\label{eq:Lambda_metrica_Ernst_Schwarzschild}
\end{equation}
\begin{equation}
\omega'=0\ ,
\label{eq:omega_metrica_Ernst_Schwarzschild}
\end{equation}
\begin{equation}
r_{+}=2M\ ,
\label{eq:sol_Delta_metrica_Ernst_Schwarzschild_1}
\end{equation}
\begin{equation}
r_{-}=0\ .
\label{eq:sol_Delta_metrica_Ernst_Schwarzschild_2}
\end{equation}

Then, from Eqs.(\ref{eq:mov_angular_Ernst})-(\ref{eq:mov_radial_Ernst}):
\begin{equation}
\frac{1}{\sin\theta}\frac{d}{d\theta}\left(\sin\theta\frac{dS}{d\theta}\right)+\left(\lambda_{lm}-\frac{m^{2}}{\sin^{2}\theta}\right)S=0\ ,
\label{eq:mov_angular_Ernst_Schwarzschild}
\end{equation}
\begin{equation}
\frac{d}{dr}\left(\Delta\frac{dR}{dr}\right)+\left[\frac{\omega^{2} r^{4}}{\Delta}-(\lambda_{lm}+B_{0}^{2}m^{2}r^{2})\right]R=0\ .
\label{eq:mov_radial_Ernst_Schwarzschild}
\end{equation}
Therefore, the general exact solution for the angular equation, $S=S(\theta)$, is
\begin{equation}
S(\theta)=C_{1}\ P_{l}^{m}(\cos\theta)+C_{2}\ Q_{l}^{m}(\cos\theta)\ ,
\label{eq:solucao_geral_angular_Ernst_Schwarzschild}
\end{equation}
where $P_{l}^{m}(\cos\theta)$, $Q_{l}^{m}(\cos\theta)$ are the associated Legendre functions with $\lambda_{lm}=l(l+1)$. As to the radial equation, the general exact solution can be written as follows
\begin{eqnarray}
R(x) & = & \mbox{e}^{\frac{1}{2}\alpha x}x^{\frac{1}{2}\beta}(x-1)^{\frac{1}{2}\gamma}\{C_{1}\ \mbox{HeunC}(\alpha,\beta,\gamma,\delta,\eta;x)\nonumber\\
&& +C_{2}\ x^{-\beta}\ \mbox{HeunC}(\alpha,-\beta,\gamma,\delta,\eta;x)\}\ ,
\label{eq:solucao_geral_radial_Ernst_Schwarzschild}
\end{eqnarray}
where $C_{1}$ and $C_{2}$ are constants, and the parameters $\alpha$, $\beta$, $\gamma$, $\delta$, and $\eta$ are given by
\begin{equation}
\alpha=4iM(\omega ^2-B_{0}^2 m^2)^{\frac{1}{2}}\ ,
\label{eq:alpha_radial_Ernst_Schwarzschild}
\end{equation}
\begin{equation}
\beta=4iM\omega\ ,
\label{eq:beta_radial_Ernst_Schwarzschild}
\end{equation}
\begin{equation}
\gamma=0\ ,
\label{eq:gamma_radial_Ernst_Schwarzschild}
\end{equation}
\begin{equation}
\delta=4M^{2}(B_{0}^{2}m^{2}-2\omega^{2})\ ,
\label{eq:delta_radial_Ernst_Schwarzschild}
\end{equation}
\begin{equation}
\eta=-(\delta+\lambda_{lm})\ .
\label{eq:eta_radial_Ernst_Schwarzschild}
\end{equation}

Thus, the resonant frequencies and the regular partial wave expression for massless scalar fields, in this background, are given by
\begin{equation}
\omega-\frac{2\omega^{2}-B_{0}^{2}m^{2}}{2(\omega^{2}-B_{0}^{2}m^{2})^{\frac{1}{2}}}=\frac{i}{2M}(n+1)\ ,\qquad n=0,1,2,\ldots\ ,
\label{eq:energy_levels_Ernst_Schwarzschild}
\end{equation}
\begin{equation}
R_{\lambda}(r) \sim C_{l}\ \frac{1}{r} \sin[k_{0,B_{0}}r-\gamma_{0,B_{0}}\ln r+\delta_{l}(\omega)]\ ,
\label{eq:sol_scattering_wave_Ernst_Schwarzschild_x_2}
\end{equation}
with
\begin{equation}
k_{0,B_{0}}^{2} \equiv \omega ^2-B_{0}^{2}m^{2}\ ,
\label{eq:k0_radial_Ernst_Schwarzschild_x}
\end{equation}
\begin{equation}
\gamma_{0,B_{0}} \equiv \frac{MB_{0}^{2}m^{2}-2M \omega ^2}{(\omega ^2-B_{0}^{2}m^{2})^{\frac{1}{2}}}\ ,
\label{eq:gamma0_radial_Ernst_Schwarzschild_x}
\end{equation}
where the phase shift $\delta_{l}(\omega)$ can be obtained by numerical calculation, as is shown for instance by Chen et al. \cite{EurPhysJC.73.2395}.

The relative scattering probability and the Hawking radiation spectrum, in this case, are given by
\begin{equation}
\Gamma_{+}=\left|\frac{\Psi_{out}(r>r_{+})}{\Psi_{out}(r<r_{+})}\right|^{2}=\mbox{e}^{-\frac{2\pi}{\kappa_{+}}\omega}\ ,
\label{eq:taxa_refl_Ernst_Schwarzschild}
\end{equation}
\begin{equation}
\left|N_{\omega}\right|^{2}=\frac{1}{\mbox{e}^{\frac{\hbar\omega}{k_{B}T_{+}}}-1}\ .
\label{eq:espectro_rad_Ernst_2_Schwarzschild}
\end{equation}
Therefore, for a Schwarzschild black hole surrounded by a magnetic field, the resulting Hawking radiation spectrum of scalar particles also has a thermal character. An interesting feature of this case is that the total energy of radiated scalar particles is not affected due to the presence of the external magnetic field.
%
%%%%%%%%%%%%%%%%%%%%%%%%%%%%%%%%%%%%%%%%%%%%%%%%%%%%%%%%%%%%%%%%%%%%%%%%%%%%%%%%%%%%%%%%%%%%%% Conclusions
%
\section{Conclusions}
We used the confluent Heun functions to investigate some processes associated with scalar fields in different background spacetimes, as the existence of resonant frequencies, the Hawking radiation and the scattering phenomenon.

First we considered the Kerr-Newman-Kasuya spacetime (dyon black hole) and examine the obtained resonant frequencies, discuss the Hawking radiation and studied the scattering of charged massive scalar fields. Due to the difficulties to obtain analytic expressions in the spacetime under consideration, we particularized to some special cases, namely, Kerr and Schwarzschild black holes, for massless scalar field and obtained analytic results which recover the well known results of the literature \cite{ZhEkspTeorFiz.64.48,PhysRevD.22.2331,ClassQuantumGrav.18.1939,PhysRevD.19.451,IntJTheorPhys.23.991,JMathPhys.17.688}.

In the case of the Reissner-Nordstr\"{o}m black hole surrounded by a magnetic field (Ernst spacetime), we obtained similar results, but now for a massless scalar field. Once again, the resonant frequencies, the Hawking radiation and the scattering process were investigated. The obtained results do not permit us to get analytic expressions. In order to overcome this difficult, we considered a special case corresponding to a Schwarzschild black hole surrounded by a magnetic field, in which case, we obtained an analytic expression for the regular partial wave, which is in accordance with the result presented in the literature \cite{EurPhysJC.73.2395}.
%
%%%%%%%%%%%%%%%%%%%%%%%%%%%%%%%%%%%%%%%%%%%%%%%%%%%%%%%%%%%%%%%%%%%%%%%%%%%%%%%%%%%%%%%%%%%%%% acknowledgments
%
\ack The authors would like to thank Conselho Nacional de Desenvolvimento Cient\'{i}fico e Tecnol\'{o}gico (CNPq) for partial financial support. H.S.V. is funded through the research Project No. 140612/2014-9. V.B.B. is partially supported through the research Project No. 304553/2010-7.
%
%%%%%%%%%%%%%%%%%%%%%%%%%%%%%%%%%%%%%%%%%%%%%%%%%%%%%%%%%%%%%%%%%%%%%%%%%%%%%%%%%%%%%%%%%%%%%% thebibliography
%

%
%%%%%%%%%%%%%%%%%%%%%%%%%%%%%%%%%%%%%%%%%%%%%%%%%%%%%%%%%%%%%%%%%%%%%%%%%%%%%%%%%%%%%%%%%%%%%%
%
\end{document}